\documentclass{ws-procs975x65}
\usepackage{graphicx}
\usepackage{latexsym}
\usepackage{amssymb}

\begin{document}

\title{Multi-accretion events from corotating and counterrotating SMBHs tori}
\author{Daniela Pugliese$^*$ and  Zdenek Stuchl\'{\i}k}

\address{Institute of Physics and Research Centre of Theoretical Physics and Astrophysics, Faculty of Philosophy\&Science, Silesian University in Opava,
 Bezru\v{c}ovo n\'{a}m\v{e}st\'{i} 13, CZ-74601 Opava, Czech Republic\\
E-mail: d.pugliese.physics@gmail.com;zdenek.stuchlik@physics.cz}

\begin{abstract}
Ringed accretion disks (\textbf{RADs})  are aggregates  of corotating and counterrotating   toroidal accretion disks  orbiting  a central Kerr super-massive Black Hole (\textbf{SMBH}) in \textbf{AGNs}.
 The dimensionless spin of the central \textbf{BH} and the fluids relative rotation  are proved to strongly affect the \textbf{RAD} dynamics.  There is   evidence of a strict correlation between \textbf{SMBH} spin, fluid rotation and magnetic fields in \textbf{RADs} formation and evolution. 
Recently, the model was extended to consider \textbf{RADs}  constituted by several magnetized accretion tori  and the   effects of a toroidal magnetic field  in \textbf{RAD} dynamics have been investigated.
The analysis poses constraints on tori formation and emergence of \textbf{RADs} instabilities in the phases of accretion onto the central attractor and tori collision emergence.
 Magnetic fields and fluids rotation are proved  to be  strongly constrained and influence tori formation and evolution in \textbf{RADs}, in dependence  on the toroidal magnetic fields parameters.
 Eventually, the \textbf{RAD} frame investigation constraints specific classes of tori  that could  be observed around some specific \textbf{SMBHs} identified by their dimensionless spin
\end{abstract}

\date{\today}
\def\be{\begin{equation}}
\def\ee{\end{equation}}
\def\bea{\begin{eqnarray}}
\def\eea{\end{eqnarray}}
\def\Sie{\mathcal{S}}
\newcommand{\bt}[1]{\mathbf{\mathtt{#1}}}
\newcommand{\tb}[1]{\textbf{{{#1}}}}
\newcommand{\rtb}[1]{\textcolor[rgb]{1.00,0.00,0.00}{\tb{#1}}}
\newcommand{\btb}[1]{\textcolor[rgb]{0.00,0.00,1.00}{\tb{#1}}}
\newcommand{\otb}[1]{\textcolor[rgb]{1.00,0.50,0.00}{\tb{#1}}}
\newcommand{\gtb}[1]{\textcolor[rgb]{0.00,.50,0.00}{\tb{#1}}}
\newcommand{\ptb}[1]{\textcolor[rgb]{0.70,0.00,0.70}{\tb{#1}}}
\newcommand{\il}{~}
\newcommand{\Qa}{\mathcal{Q}}

\bodymatter


\section{Introduction}
Ringed accretion disks (\textbf{RADs}) are aggregates of axi-symmetric tori orbiting Kerr  \textbf{SMBHs}. The orbiting agglomerate is composed by both corotating and counterrotating tori. These configurations  might be generated after different periods of the attractor life, where the angular momentum of the material in accretion on the central attractor in different phases of the \textbf{BH} life could have very different direction and magnitude.
The \textbf{\textbf{RAD}} model  was first developed considering tori centered on the equatorial plane of the central \textbf{BH}.
\begin{figure}
  \begin{center}
  \includegraphics[width=4cm]{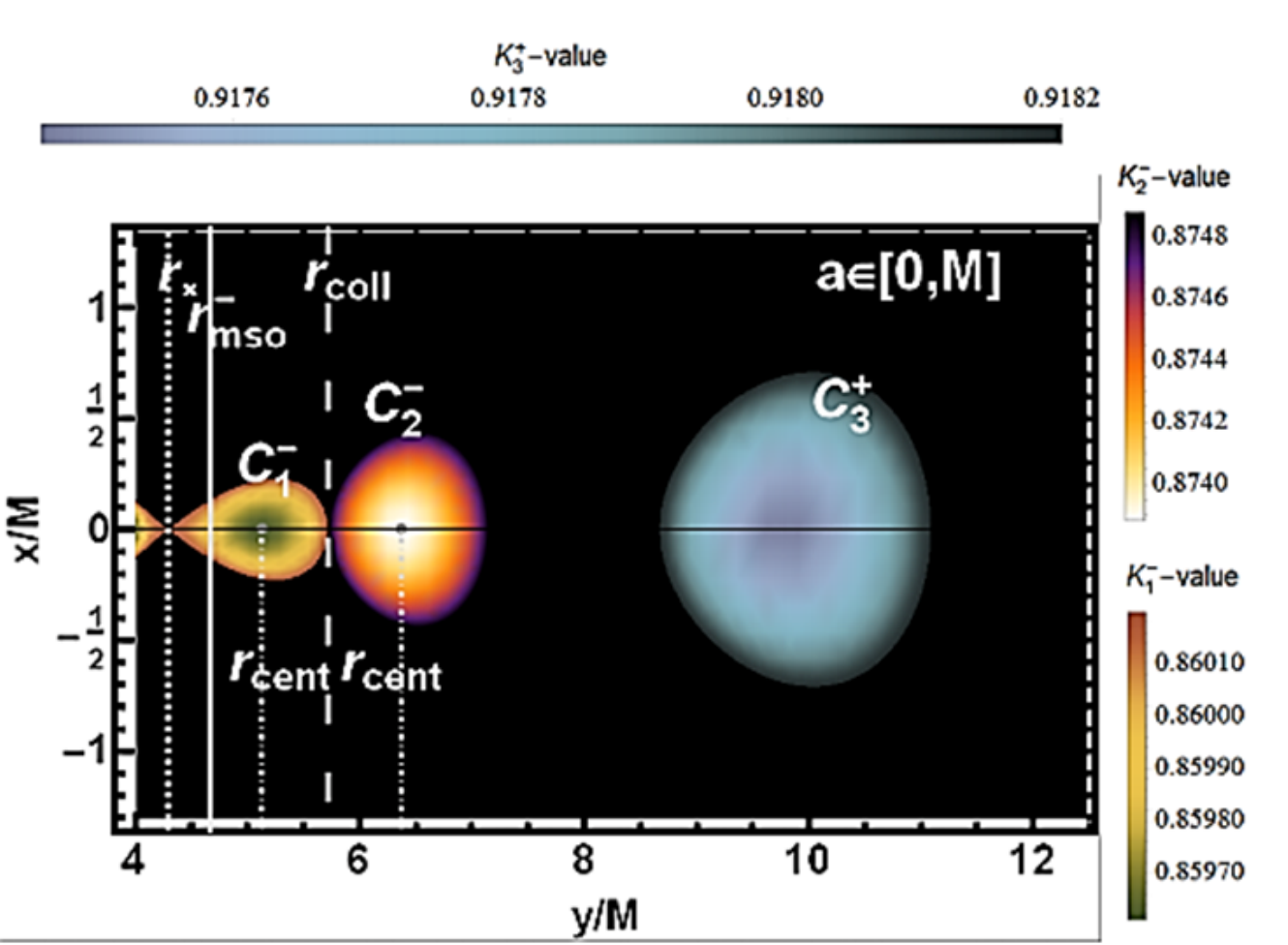}
   \includegraphics[width=4cm]{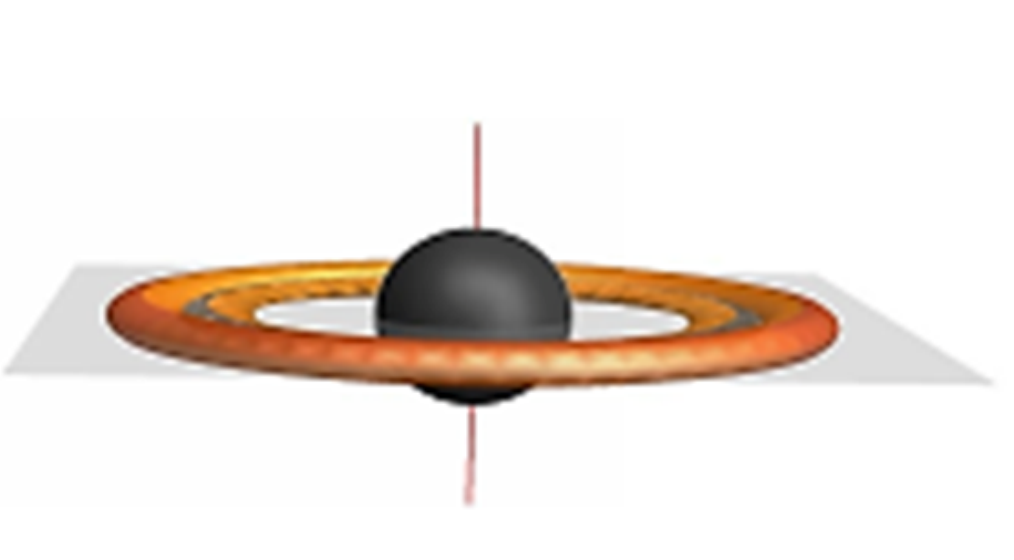}
   \includegraphics[width=9cm]{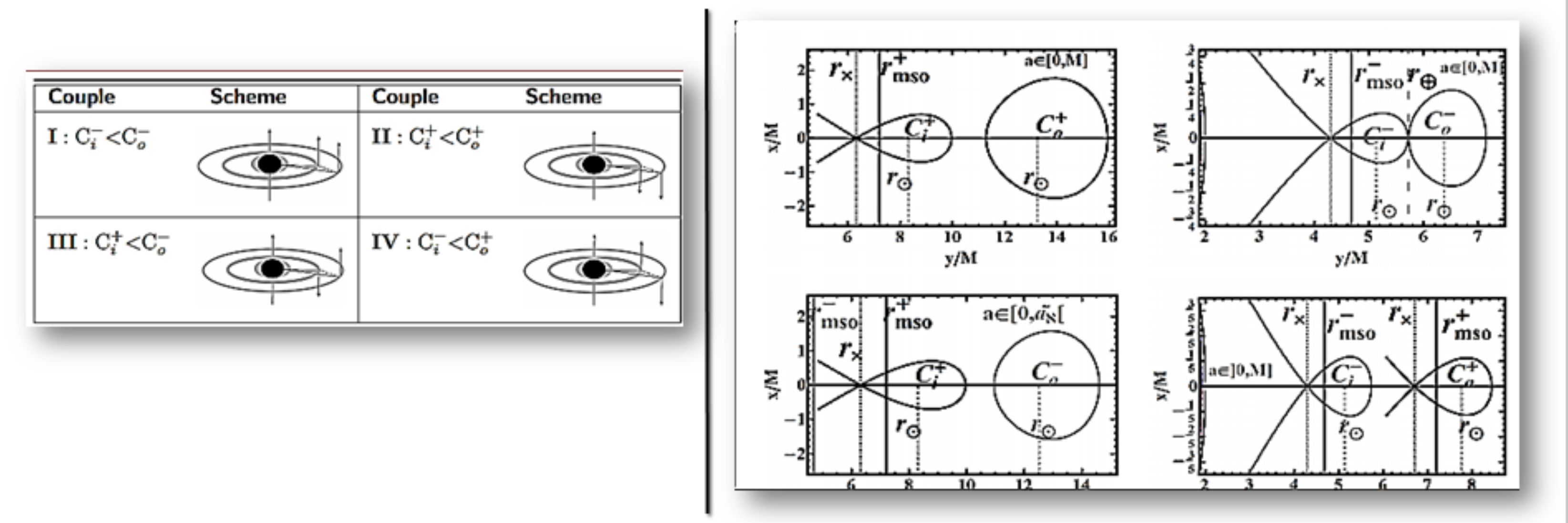}
  \caption{Upper Left: \textbf{\textbf{RAD}} of the order 3, 2D-GRHD density profiles. Upper Right: \textbf{\textbf{RAD}} of the order 2, 3D-GRHD density profiles of corotating colliding tori. Below Left: schemes for the   $\ell$corotating ($\ell$ \textbf{c}) or $\ell$counterrotating ($\ell$ \textbf{r}) couples. Below  Right: correspondent \textbf{\textbf{RAD}} density profiles. Upper line is for the  ($\ell$ \textbf{c}) of counterrotating (left) and corotating (right) tori\cite{multy,long}.}\label{Fig:3coll}
  \end{center}
\end{figure}
\textbf{RAD} can be associated to jet emission in many ways, because of  the inner edge of accreting disks jet launch point correlation and because  the model adopted for each toroidal disk component of the aggregate provides an open solution, proto-jets, consisting of shells funnels of matter along the axis of the \textbf{BH}--\cite{proto-jet,open,long}. 
It should be noted  that the occurrence  of double accretion in the  \textbf{RAD} is restricted to the case of two tori only, for a couple  made by an inner corotating torus and by an outer counterrotating one.
Consequently, the double jet shell emission  in this system  is limited to a jet from  external counterrotating matter and an inner jet from the internal corotating matter-
  \cite{proto-jet}. It was also  provided a general classification of the \textbf{RAD} and of the attractors according to their dimensional spin, providing an indication on the configurations and situation where observational evidences can be found. In  \cite{ringed} a large part of modellization was dedicated to  describing the \textbf{RAD} as a single disk: in fact it has been shown that such kind od agglomerate is  generically a geometrically thin disk, axis-symmetric,  with knobby surface, and a very varied internal activity--Figs\il(\ref{Fig:3coll}). The \textbf{RAD} could be disguised as one disk characterized buy  articulated phases of super-Eddington accretion, which could explain  the masses problem  of super-massive \textbf{BH} at high redshift, after a series of interrupted phase of accretion--\cite{multy}.

 The model was introduced in \cite{Evolutive} and fully developed in \cite{ringed}. In \cite{open} the unstable configurations were analyzed and particularly constrains on locations of inner and outer edges of the \textbf{\textbf{RAD}} and the toroidal components  were given. A detail analysis of \textbf{RAD} tori sequences as remnants of multiple accreting periods of Kerr \textbf{SMBHs}, in active galactic nuclei (\textbf{AGNs}) was considered in \cite{long}. In \cite{dsystem}, the constraints on double configurations, constituted by two tori centered on the \textbf{BH} were given considering the possibility of collision emergence.
Energetics associated to the \textbf{RAD} processes occurring in the ringed structure between its components and particularly the evaluation of  \textbf{BH} accretion rates is  presented in \cite{multy}.
Here we present the main results of the hydrodynamic (HD) and magnetized \textbf{RAD} based on the investigation in  \cite{ringed,open,dsystem,long,jet,multy,Fi-Ringed}.

\textbf{Overview}
In Sec.\il(\ref{Sec:Stationar})  General considerations on hydrodynamic \textbf{\textbf{RAD}} models are discussed.
In Sec.\il(\ref{Sec:constr})
constraints are discussed.
In Sec.\il(\ref{Sec:influence}) we discuss  the
effects of  toroidal magnetic fields  on \textbf{RADs}.
Concluding remarks follow in Sec.\il(\ref{Sec:remark}.
 \section{Orbiting Axi-symmetric tori in a Kerr spacetime
}\label{Sec:Stationar}
Because of the stationarity (time  $t$ independence) and the axis-symmetry  ($\phi$-independence) of each toroid of the aggregate, each torus is regulated by the Euler equation only
with  a barotropic equation of state $p=p(\rho)$, there is
\begin{equation}\label{E:II}
 	\int^p_z\frac{d p}{p +\rho}=W(p)-W(0)=-\ln\frac{U_t}{(U_t)_{in}}+\int^l_{l_in}\frac{\Omega d l}{1-\Omega l},
\end{equation}
where  $\Omega$ is the fluid relativistic angular frequency,
 $\ell$  specific angular momenta,  here assumed  constant  and conserved inside each torus but not in the \textbf{\textbf{RAD}}, $(t, r, \phi, \theta)$ are  Boyer-Lindquist  (B-L)
 coordinates.  We consider $\ell>0$
for \emph{corotating}   and $\ell<0$  for  \emph{counterrotating} fluids, within  the notation $(\mp)$   respectively.
$a=J/M\in]0,M]$  is the  \textbf{BH} spin, $J$ is the
total angular momentum of the gravitational source and $M$ is the  gravitational mass parameter.
For the tori couple $(C_{(i)}, C_{(o)})$,  orbiting  in   the equatorial plane of a given Kerr \textbf{BH}  with specific angular momentum $(\ell_{(i)}, \ell_{(o)})$,   we  introduce   the concept  of
 \emph{$\ell$corotating} (\textbf{$\ell$c}) disks,  defined by  the condition $\ell_{(i)}\ell_{(o)}>0$, and \emph{$\ell$counterrotating} (\textbf{$\ell$r}) disks defined  by the relations   $\ell_{(i)}\ell_{(o)}<0$.  The two (\textbf{$\ell$c}) tori  can be both corotating, $\ell a>0$, or counterrotating,  $\ell a<0$, with respect to the central attractor $a>0$.
Each  \textbf{\textbf{RAD}} torus  is a General Relativistic  (GR)   model  based  on the Boyer theory of the equi-pressure surfaces.
Boyer surfaces are  constant pressure  surfaces and  $\Sigma_i=\Sigma_{j}$ for \({i, j}\in(p,\rho, \ell, \Omega) \). Toroidal surfaces correspond to the   equipotential surfaces, critical points of $V_{eff}(\ell) $ as function of $r$, thus  solutions of   $W:\;  \ln(V_{eff})=\rm{c}=\rm{constant}$ or $V_{eff}=K=$constant
{$\mathbf{C}$}--cross sections of the{ closed} Boyer surfaces (equilibrium quiescent torus);
{$\mathbf{ C_x}$ }--cross sections of the {closed cusped}  Boyer surfaces (accreting torus);
{$\mathbf{O_x}$}--cross sections of the {open cusped}  Boyer surfaces, generally associated to proto-jet configurations. In the following we use the notation $()$ to indicate a configuration which can be closed, $C$,  or open $O$--Figs\il(\ref{Fig:3coll}).
The model in constructed  investigating the angular momentum distribution inside the disk (which is not constant):
\be\label{Eq:laud}\pm\ell^{\mp}_n=\pm\left.\frac{a^3+a r_n (3 r_n-4M)\mp\sqrt{r_n^3 \Delta_n^2}}{a^2-(r_n-2M)^2 r_n}\right|_{r_n^{\ast}},\quad \Delta_n\equiv r_n^2-2 M r_n+a^2\ee
where $n$  is  for the toroidal \textbf{\textbf{RAD}}  component, and  $\Delta_n$  is here a metric factor.
In fact an essential part of the \textbf{\textbf{RAD}} analysis is the characterization of the boundary conditions on each torus in the agglomerate and  of the disk. We considered also
the function
\bea\mathrm{K}^{\pm}(\Delta_{(\pm)},r)\equiv\left.\frac{1}{2} \sqrt{\frac{r \left[(\Delta_--\Delta_+)^2+4 (r-2) r\right]}{2 \Delta_-^2-\Delta_- \Delta_+ r+r^3}}\right|_{\bar{\ell}^{\mp}_n},\quad\ell=\frac{\Delta_-+\Delta_+}{2},\; a=\frac{(\Delta_+-\Delta_-)}{2},\eea
(dimensionless units),  provides constrains on the  matter density distribution inside the disk and stitching together the \textbf{\textbf{RAD}} tori.
\section{Constraints}\label{Sec:constr}
%
In the  Bondi quasi spherical  accretion, the fluid angular momentum is  everywhere  smaller  than  the  Keplerian  one and
therefore   dynamically  unimportant. In this analysis, however, we consider a full GR model for each \textbf{RAD} toroid where in fact there exists
an   extended region where the fluids angular momentum in the torus  is larger (in magnitude) than the Keplerian (test particle) angular momentum. More precisely as first canvas model we adopt   for each toroid a thick, opaque (high optical depth) and super-Eddington, radiation pressure supported  accretion disk (in the toroidal disks, pressure gradients are crucial) cooled by advection with low viscosity. 
As a consequence, during the evolution of dynamical processes, the functional form of the angular
momentum and entropy distribution depends on the initial conditions of the system and on
the details of the dissipative processes.
From these considerations, using the distribution of relativistic  specific angular momentum   in the  \textbf{\textbf{RAD}} as in  Eq.\il(\ref{Eq:laud}) we
can fix, as in  \cite{ringed,open,dsystem,multy}, the  constraints on the range of variation of the inner edge of accreting  torus,   $r_x$, and on the point of maximum density  (pressure) in each torus, $r_{cent}$, in dependence from the range of variation of the specific angular momentum in the disk.
Precisely, constraints  on the angular momentum $\ell$ ranges are as follow:

\medskip

{\small\textbf{[Range- L1]}
$
\mp \mathbf{L1}^{\pm}\equiv[\mp \ell_{mso}^{\pm},\mp\ell_{mbo}^{\pm}[$  where   topologies $(C_1, C_{\times})$ are possible,  with accretion point in  $r_{\times}\in]r_{mbo},r_{mso}]$ and center with maximum pressure  $r_{cent}\in]r_{mso},\rho_{mbo}]$;
\\
\textbf{[Range- L2]}
$\mp \mathbf{L2}^{\pm}\equiv[\mp \ell_{mbo}^{\pm},\mp\ell_{\gamma}^{\pm}[ $ where   topologies    $(C_2, O_{\times})$ are possible,  with unstable point  $r_{j}\in]r_{\gamma},r_{mbo}]$  and  center with maximum pressure $r_{cent}\in]\rho_{mbo},\rho_{\gamma}]$;
\\
\textbf{[Range- L3]}
$\mp \mathbf{L3}^{\pm}\equiv\ \ell \geq\mp\ell_{\gamma}^{\pm} $    where only equilibrium torus  $C_3$  is possible with center $r_{cent}>\rho_{\gamma}$;}
(in the following  indices $i\in\{1,2,3\}$ refer to the  ranges of angular momentum $\ell\in \mathbf{Li}$)
 being $mso=$marginally stable orbit and $mbo=$marginally bounded orbit, $\gamma=$marginally circular orbit (photon orbit.)
Alongside the geodesic structure of the Kerr spacetime represented by the set of radii $R\equiv (r_{mso}^{\pm}, r_{mbo}^{\pm},r_{{\mathrm{\gamma}}}^{\pm})$, and $r_{\mathcal{M}}^{\pm}$  solution  of  $\partial_r^2\ell=0$, relevant to the location of the disk center and outer edge are $R_{\rho}\equiv (\rho_{mbo}^{\pm}, \rho_{{\mathrm{\gamma}}}^{\pm},\rho_{\mathcal{M}}^{\pm}):$:
\bea&&\nonumber
r_{\mathrm{\gamma}}^{\pm}<r_{mbo}^{\pm}<r_{mso}^{\pm}<
 \rho_{\mathrm{mbo}}^{\pm}<
 \rho_{{\mathrm{\gamma}}}^{\pm}\quad\mbox{where}\quad \rho_{\mathrm{mbo}}^{\pm}:\;\ell_{\pm}(r_{mbo}^{\pm})=
 \ell_{\pm}(\rho_{\mathrm{mbo}}^{\pm})\equiv \mathbf{\ell_{mbo}^{\pm}},
\\&&\label{Eq:conveng-defini}
  \rho_{{\mathrm{\gamma}}}^{\pm}: \ell_{\pm}(r_{{\mathrm{\gamma}}}^{\pm})=
  \ell_{\pm}(\rho_{{\mathrm{\gamma}}}^{\pm})\equiv \mathbf{\ell_{{\mathrm{\gamma}}}^{\pm}},
 \quad
\rho_{\mathcal{M}}^{\pm}: \ell_{\pm}(\rho_{\mathcal{M}}^{\pm})= \ell_{\mathcal{M}}^{\pm}.
\eea
This expanded structure rules good  part of the geometrically  thick disk physics and multiple structures. The presence of  these radii stands as one of the main effects of the  presence of a strong curvature  of  the background geometry\cite{multy,open,long}.
\subsection{The RAD constraints }
The \textbf{RAD} of the order $n=2$ (the order  $n$ is the number of toroidal components)
can be composed by the the following toridal couples:

{\textbf{{i)}}} $C_{\times}^{\pm}< C^{\pm}$,
{{\textbf{ii)}}}  $C_{\times}^{+}< C^{\pm}$, {{\textbf{iii)}}} $C_{\times}^{-}< C^{\pm}$ and
{\textbf{{iv)}}} $C_{\times}^{-}< C_{\times}^{+}$.
We indicated with $<$   the relative rotation of maximum density points  in the tori--Figs\il(\ref{Fig:3coll}).
The situation concerning the  emergence of more
accretion points  in \textbf{\textbf{RAD}} and presence of double accretion and screening tori, is summarized as follows:
{\small
 \bea
\mbox{for}\quad a\in]0,M[:\quad
 \underset{\textbf{(a)}}{\underbrace{\mathbf{C_x^-}}}<\underset{\textbf{(b)}}{\underbrace{...<C_-<...}}
 <\underset{\textbf{(c)}}{\underbrace{\mathbf{C_x^+}}}<...\underset{\textbf{(d)}}{\underbrace{<C_{\pm}<}}...
 \eea}
where
$\mathbf{(a)}$ is the inner ring in accretion, $\mathbf{(b)}$ is the  inner subsequence of corotating tori in equilibrium, $\mathbf{(c)}$ is  the outer accreting tori  and  $\mathbf{(d)}$ is the outer (mixed or separated) subsequence composed by equilibrium tori only. Further constraints (for the specific angular momentum, elongation and number $n$ of tori) on the subsequences  $\mathbf{(b)}$ and $\mathbf{(d)}$ depend on the attractor spin mass ratio--Figs\il(\ref{Fig:3coll}).
More precisely:

\textbf{The \textbf{\textbf{RAD}} canvas}

\medskip

{\small
$\bullet$
There is a maximum $n=2$   of  accreting tori of the king $C_x^-<C_x^+$ around supermassive   \textbf{BHs} with $a\neq0$. Such tori are strongly   constrained in their values of  $\ell$ and $K$ depending on the spin-mass ratio of the attractor.
\\
 $\bullet$``Screening''-tori  located between the accreting disk and the central \textbf{BH} in a \textbf{RAD} sequence as $C_x^-<C^-<...<C_x^+<C^{\pm}$ are possible only as corotating, quiescent  $C^-$  inner tori.
  A  screening torus  is  a corotating (non-accreting) torus eventually detectable by  X-ray spectra emission obscuration.
\\
 $\bullet$
If a counterrotating torus of a \textbf{RAD} is accreting onto the central \textbf{BH}, then a \textbf{\textbf{RAD}} with a corotating outer  $C^-_1$ torus,  is as
$()_x^-<C_x^+<C_1^-<C^{\pm}$,  only orbiting    ``slow'' \textbf{SMBHs} ($a<0.46M$).
\\
\textbf{[$\bullet$-\textbf{{Corotating  Tori}}]}
 A  corotating torus can be the outer  of a couple with an   inner counterrotating accreting torus. Then the outer torus may be corotating (non  accreting), or counterrotating in accretion or  quiescent.    Both  the inner corotating  and the outer counterrotating  torus of the couple  can accrete onto the attractor.
\\
\textbf{[$\bullet$-\textbf{{Counterrotating  Tori}}]}
A counterrotating  torus can therefore reach the instability being the inner of a (\textbf{$\ell$r})-(\textbf{$\ell$c}), or the outer torus of a (\textbf{$\ell$r})  couple.
If  the \emph{accreting} torus is \emph{counterrotating}, $C_{\times}^+$, with respect to the Kerr attractor,  there is \emph{no} inner  counterrotating torus, but there is  $C_{\times}^+<C^{\pm}$ or $()^-<C_{\times}^+$.
}
\section{Influence of toroidal magnetic field in multi-accreting tori}\label{Sec:influence}
\begin{figure}
\begin{center}
  \includegraphics[width=7cm]{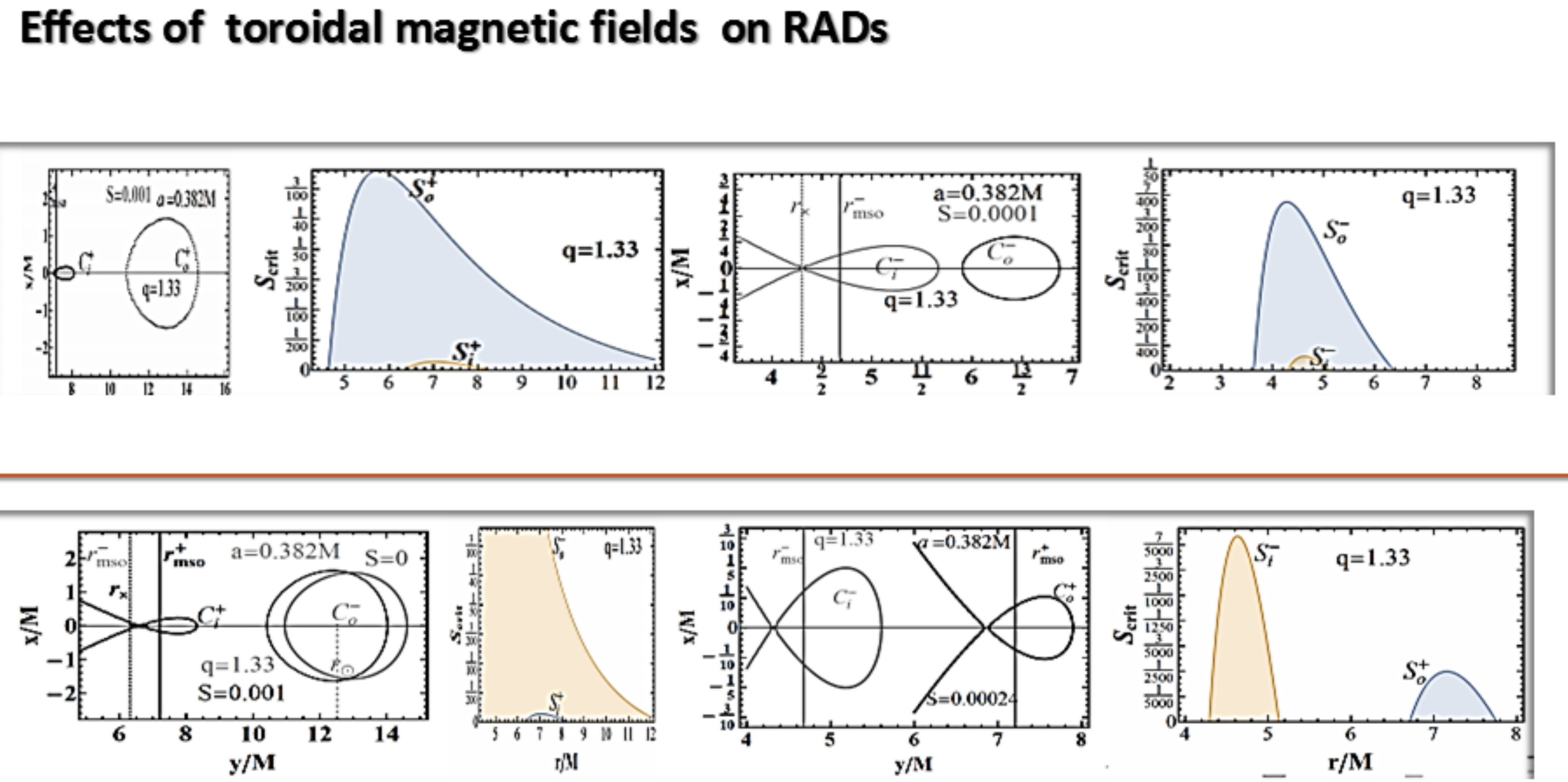}
  \caption{Magnetized \textbf{RAD} of the {order} $n=2$. Density profiles versus $\Sie$ functions profiles.
  Upper line features the $\ell$corotating couples of corotating (-) and counterrotating (+) tori.
  Bottom line features the $\ell$counterrtating couples made by $C_+<C_-$ (left)  or  $C_-<C_+$ (right). From \cite{Fi-Ringed}.}\label{Fig:slites}
  \end{center}
\end{figure}

We considered a toroidal magnetic field contribution in each  \textbf{\textbf{RAD}} component where the magnetic field is--\cite{Komissarov}:
\bea\label{RSC}
B^{\phi }=\sqrt{\frac{2 p_B}{g_{\phi \phi }+2 \ell  g_{t\phi}+\ell ^2g_{tt}}}\quad\mbox{with}\quad   p_B=\mathcal{M} \left(g_{t \phi }g_{t \phi }-g_{{tt}}g_{\phi \phi }\right){}^{q-1}\omega^q
\eea
the magnetic pressure,
$\omega$ is the fluid enthalpy, $q$  and $\mathcal{M}$ (magnitude) are constant;
$V_{eff}$ is a function of the metric and the angular momentum $\ell$.
Euler  equation  (\ref{E:II}) is modified by the term:
{\small
\bea\label{Eq:Kerr-case}
&&\partial_{\mu}\tilde{W}=\partial_{\mu}\left[\ln V_{eff}+ \mathcal{G}\right]\, \quad\mathcal{G}(r,\theta)=\Sie\left(g_{{t\phi }} g_{t\phi}-g_{tt} g_{\phi \phi}\right)^{q-1},\;\Sie\equiv\frac{q \mathcal{M} \omega^{q-1}}{q-1},
\eea
}
({where}
$a\neq0$) $q=1$, is a singular values for the magnetic parameter $\Sie$, which is negative at $q<1$, in this case excretion  tori are possible. We here concentrate on $q>1$.  
%
 We therefore consider  the equation for the
\(
\tilde{W}=K
\). 
For $\Sie=0$  (or $\mathcal{M}=0$) this  reduces to the HD case.
The \textbf{\textbf{RAD}} angular momentum distribution is:
{\footnotesize
\bea\label{Eq:dilde-f}
&&\widetilde{\ell}^{\mp}\equiv\frac{a^3+a r \left[4 \Qa (r-M) \Sie \Delta^{\Qa}+3 r-4\right]\mp\sqrt{r^3\Delta^{2} \left[1+[2\Qa (r-1)^2 r \Sie\Delta ^{\Qa-1}(2\Qa\Sie\Delta ^{\Qa}+1)]\right]}}{
\Delta^{-1}[a^4-a^2 (r-3) (r-2) r-(r-2) r \left[2 \Qa (r-1) \Sie \Delta ^{\Qa+1}+(r-2)^2 r\right]]}
%
\eea}
$\Qa\equiv q-1$ (dimensionless units). 
However  the introduction of  a toroidal magnetic field $B$, makes the study of the momentum distribution within the disk rather complicated, instead   In \cite{Fi-Ringed}  it was adopted  the function derived from
 $\Sie$-\textbf{\textbf{RAD}} parameter:
{\small
\bea\label{Eq:Sie-crit}
\mathcal{\Sie}_{crit}\equiv-\frac{\Delta^{-\Qa}}{\Qa}\frac{a^2 (a-\ell)^2+2 r^2 (a-\ell) (a-2 \ell)-4 r (a-\ell)^2-\ell^2 r^3+r^4}{2  r  (r-1)\left[r (a^2-\ell^2)+2 (a-\ell)^2+r^3\right]}
\eea}
(dimensionless units)
capable of setting  the  location of maximum density points   in the disk and the existence and location  of  the instability points.
Eq.\il(\ref{Eq:Sie-crit}) clearly enucleates the  magnetic field contribution in the   $\Qa$ term, while interestingly highlights the role of the parameters  $\ell$ versus $a$--\cite{proto-jet,Evolutive}.
The effects of the toroidal magnetic field in the \textbf{\textbf{RAD}} composition are evident in Fig.\il(\ref{Fig:slites}).
In \cite{Fi-Ringed}  it is noted that a \textbf{RAD} can be formed by having generally a sufficiently small parameter  $(\Sie q)$.
Profiles of  ($\ell$\textbf{c}) cases are similar independently by the corotation or counterrotation of the fluids in the \textbf{RAD} with the respect to the central Kerr \textbf{SMBH}. Generally the inner torus has maximum values of the  $\Sie$ smaller the maximum $\Sie$ found in the outer tori.
The most interesting results perhaps  emerge in the case of ($\ell$\textbf{r}) couples  where it is clear that  the magnetic profiles for the  couple  $C_-<C_+$  (where double accretion occurs) are  radically different from the case $C_+<C_-$. The analysis shows also the importance of the  coupling between the toroidal  component of the magnetic field and the fluid angular momentum, particularly in the counter-rotating case, $\ell<0$; for this case, for values  $q<1$ excretion can arise\cite{Fi-Ringed}--\cite{excre1}. 
\section{Concluding Remarks and Future perspectives}\label{Sec:remark}
The \textbf{RAD} investigation  has  revealed   a very rich scenario  with interesting phenomenology linked to the activity inside the agglomeration, correlated  to the processes of instability and interaction between tori and the tori and central  \textbf{SMBH} attractor.
In this scenario constrains on the
   presence of  screening  and obscuring  tori were discussed. The analysis of the screening and obscuring tori in particular  could lead to observational evidences of a double tori \textbf{RAD} system  from  the emission spectrum as  X-ray emission screening, showing as   fingerprints of the discrete radial profile of the \textbf{RAD}.
  In \cite{S11etal} then  relatively indistinct excesses
 of the relativistically broadened  emission-line components were predicted arising in a well-confined
radial distance in the accretion structure
  originating by a series of  episodic accretion events.
 Another  interesting aspect of this model is the possibility of having inter-disk activity resolved in tori collision  or double accretion phase with a double jet emission phase.
From the  \textbf{RAD} investigation this activity is   limited to only two specific tori of aggregate made by a special couple constituted by a corotating and a counterrotating torus.
As sideline result we provided a full characterization of the counterrotating tori in the multi-accreting systems.
 This  model is designed for an extension to a dynamic GRMHD setup.
Currently the  toroidal model adopted  to picture  each \textbf{RAD} components is   used as initial configuration for such systems.
Another significant aspect is the possibility of inter-correlate the oscillations of the \textbf{RAD} components  with  \textbf{QPOs}. 
Generally instabilities of
such configurations, can reveal to be of great  significance for the High Energy Astrophysics related
to accretion onto supermassive \textbf{BHs}, in Quasars and \textbf{AGNs}.
Such activities could be targeted  by the planned X-ray observatory \texttt{\textbf{ATHENA}}\footnote{\url{http://the-athena-x-ray-observatory.eu/}}.


\subsubsection*{Acknowledgments}
D.P. acknowledges support from the Junior GACR grant of the Czech Science Foundation No:16-03564Y.

\end{document}